\documentclass[ aps,%
floatfix,%
final,%
notitlepage,%
oneside,%
onecolumn,%
nobibnotes,%
nofootinbib,%
superscriptaddress,%
showpacs,%
]%
{revtex4}

\usepackage{epsfig}
\usepackage{amsfonts}
\usepackage{amsmath}
\usepackage{epsfig}
\usepackage{graphics}

\newcommand{\beq}{\begin{eqnarray}}\newcommand{\eeq}{\end{eqnarray}}
\newcommand{\beqa}{\begin{eqnarray*}}\newcommand{\eeqa}{\end{eqnarray*}}

\begin{document}

\title{Mass of nonrelativistic meson from leading twist distribution amplitudes. }
\author{V.V. Braguta}

\email{braguta@mail.ru}

\affiliation{Institute for High Energy Physics, Protvino, Russia}

\begin{abstract}
In this paper distribution amplitudes of pseudoscalar and vector nonrelativistic mesons
are considered. Using equations of motion for the distribution amplitudes, it is derived 
relations which allow one to calculate the masses of nonrelativistic pseudoscalar 
and vector meson if the leading twist distribution amplitudes are known.  
These relations can be also rewritten as relations between the masses
of nonrelativistic mesons and infinite series of QCD operators, 
what can be considered as an exact version of Gremm-Kapustin relation in NRQCD.
\end{abstract}


\maketitle

\newcommand{\ins}[1]{\underline{#1}}
\newcommand{\subs}[2]{\underline{#2}}
%%%%%%%%%%%%%%%%%%%%%%%%%%%%%%%%%%%%%%%%%%%%%%%
\vspace*{-1.cm}
\section{Introduction}

Study of hard exclusive processes can lead to a better understanding of 
mechanisms of hadrons production and hardrons structure. 
The description of such processes is based on the factorization 
theorem \cite{Lepage:1980fj, Chernyak:1983ej}. Within this theorem the 
amplitude of hard exclusive process can be separated into two parts. 
The first part is partons production at very small distances, which can 
be treated within perturbative QCD. The second part is 
hardronization of the partons at larger distances. This part contains  information about 
nonperturbative dynamic of strong interaction. For hard exclusive processes it can be 
parameterized by process independent distribution amplitudes (DA), which can be 
considered as hadrons' wave functions
at light-like separation between the partons in the hadron. 

Recently, the leading twist (twist-2) DAs of nonrelativistic mesons have become the object of intensive 
study  \cite{Bodwin:2006dm, Ma:2006hc, Braguta:2006wr, Braguta:2007fh, Braguta:2007tq, 
Choi:2007ze, Feldmann:2007id, Bell:2008er, Braguta:2008qe, Hwang:2008qi, Hwang:2009cu}. 
In papers \cite{Ball:1998je, Ball:1998sk} it was derived relations between twist-2 
and twist-3 DAs of light mesons. Unfortunately, these relations contain
DAs of higher Fock states, for which one has very poor knowledge. 
For nonrelativistic mesons, there states are suppressed by higher powers 
of relative velocoty of quark-antiqaurk pair inside meson \cite{Bodwin:1994jh}, what leads
to simplification of these relations. In papers \cite{Braguta:2008tg, Braguta:2009xu} it was shown 
that if one ignores the contribution of higher Fock states, one can write 
the solution of equations of motion for twist-3 DAs as an infinite series in Gegenbauer 
polynomials. Actually, using equations of motion one 
can find exact formula for the twist-3 DAs of nonrelativistic mesons through the 
leading twist DAs. In this paper we apply this exact formula and derive a relation
between the masses of pseudoscalar 
and vector nonrelativistic mesons and  leading twist DAs of these mesons. 

This paper is organized as follows. Next section is devoted to the derivation 
of the relation between mass and the leading twist DA of pseudoscalar nonrelativistic 
meson. In section three similar relation will be derived for the vector 
meson. 

\section{Pseudoscalar nonrelativistic mesons.}

In this section the relation which allows one find the 
mass of pseudoscalar nonrelativistic meson if the leading twist DA of this meson is known 
will be derived. To do this one needs the following 
DAs
\beq
\langle P(p) | \bar Q(z) \gamma_{\alpha} \gamma_5 [z,-z] Q(-z)| 0 \rangle_{\mu} 
= i f_P p_{\alpha} \int_{-1}^{1} d \xi e^{i(pz)} \varphi_A (\xi, \mu), \nonumber \\
\langle P(p) | \bar Q(z)  \gamma_5 [z,-z] Q(-z)| 0 \rangle_{\mu} 
= i f_P \frac {M_P^2} {2 M_Q(\mu)} \int_{-1}^{1} d \xi e^{i(pz)} \varphi_P (\xi, \mu), 
\eeq
where $\mu$ is the scale at which the axial and pseudoscalar DAs $\varphi_A(\xi, \mu), \varphi_P(\xi, \mu)$ are 
defined, $M_Q(\mu)$ is the running mass of the heavy quark $Q$, $M_P$ is the mass of the 
pseudoscalar meson $P$, $\xi$ is the fraction of the whole momentum of the meson $P$ 
carried by quark-antiquark pair, $[z,-z]$ is a gluon string which makes the corresponding 
operators gauge invariant. 

Now few comments are in order. The meson $P$ is assumed to be nonrelativistic. 
The width of  DAs of nonrelativistic meson is of order of $\xi^2 \sim v^2$, where $v$ is relative 
velocity of quark-antiquark pair in the meson $P$ \cite{Braguta:2006wr}. In the end point region 
$|\xi| \sim 1$ the motion of quark-antiquark pair is relativistic, so 
DAs in these regions must be strongly suppressed. One can even assume that 
DAs of nonrelativistic meson are zero in the end point region. However, as  was 
shown in paper \cite{Braguta:2006wr} radiative corrections, which can be taken into  
account through the evolution equation for the DAs, lead to the nonzero effect of the DAs 
in the end point region. In this paper {\it all DAs and all scale dependent quantities 
will be taken at the scale at which radiative corrections are not very important } 
and below explicit scale dependence will be omitted.
For instance, in paper \cite{Braguta:2006wr} it was shown that for charmonia mesons 
the effect of the evolution of the DAs is negligible at scale $\mu \sim \mu_0=1.2$ GeV. 
It should be noted that at this scale 
$\mu \sim \mu_0$ the DAs $\varphi_A(\xi, \mu), \varphi_P(\xi, \mu)$ are equal to each other 
at the leading order approximation in relative velocity \cite{Braguta:2006wr}.

In paper \cite{Ball:1998je} it was shown that applying equations of motion one 
can derive the following relations between the moments of the DAs 
$\varphi_A(\xi), \varphi_P(\xi)$
\beq
\langle \xi^n \rangle_P = \delta_{n0} + \frac {n-1} {n+1} \biggl ( \langle \xi^{n-2} \rangle_P 
- r \langle \xi^{n-2} \rangle_A
\biggr ),
\label{eom1}
\eeq
where $r=4 M^2_Q/M_{\eta}^2$, the moments $\langle \xi^{n} \rangle_{A,P}$ are defined as
\beq
\langle \xi^{n} \rangle_{A,P} = \int_{-1}^{1} d \xi~ \xi^n \varphi_{A,P}(\xi).
\eeq
It should be noted that actually equations (\ref{eom1}) contan also terms
which describe the contribution of higher Fock state quark-antiquark-gluon. 
However, for the nonrelativistic mesons such states are suppressed by 
large power of small relative velocity $v$ of quark-antiquark pair \cite{Bodwin:1994jh}. 
For this reason in equations (\ref{eom1}) and below such terms are disregarded. 

Equations (\ref{eom1}) can be easily solved. The solution is given 
by the following expression
\beq
\varphi_P(\xi) = C + r \int_{-1}^{\xi} \frac {\varphi'_A(t)} {1-t^2} dt, 
\label{phiP}
\eeq
where the constant $C$ can be determined from the normalization condition 
for the function $\varphi_P(\xi)$
\beq
\int_{-1}^{1} d \xi~ \varphi_P(\xi) = 1 = 2 C + r \int_{-1}^{1} d \xi  \frac {1+\xi^2} {(1-\xi^2)^2} \varphi_A(\xi).
\label{norm}
\eeq
Note that equation (\ref{phiP}) unambiguously determines the twist-3 DA $\varphi_P(\xi)$ through 
the twist-2 DA $\varphi_A(\xi)$. Actually this can always be done if one disregards 
the contribution of higher Fock states. One more example which demonstrates the last 
statement will be given below for vector nonrelativisitc meson. 

As was explained above the DA $\varphi_P(\xi)$ is zero when $\xi$ is sufficiently 
near to the end points $\xi \to \pm 1$. In order to obtain this behaviour 
when $\xi \to \pm 1$ for the function $\varphi_P(\xi)$ in equation (\ref{phiP})
one must set $C=0$, since the integral in the right hand side of 
equation (\ref{phiP}) is zero for $\xi \sim \pm 1$. If one puts 
$C=0$ to equation (\ref{norm}), one gets the formula 
\beq
\frac {M_P^2} {4 M_Q^2} = \int_{-1}^{1} d \xi  \frac {1+\xi^2} {(1-\xi^2)^2}~ \varphi_A(\xi). 
\label{massP}
\eeq
The running quark mass $M_Q$ here is taken at the scale at which radiative correction can be 
disregarded. Formula (\ref{massP}) is one of the main results of this paper.

It is known that the moments $\langle \xi^n \rangle_A$ can be expressed through the QCD operators
\beq
\langle P(p) | \bar Q \gamma_{\nu} \gamma_5 (i z^{\sigma}  {\overset {\leftrightarrow} {D}}_{\sigma} )^n Q| 0 \rangle = i
f_{P} p_{\nu} (zp)^n \langle \xi^n \rangle_A.
\label{xin}
\eeq
Here 
\beq
{\overset {\leftrightarrow} {D}}={\overset {\rightarrow} {D}}-{\overset {\leftarrow} {D}}, ~~~ {\overset {\rightarrow} {D}} = {\overset {\rightarrow} {\partial}}- 
i g B^a (\lambda^a /2).
\eeq
From this perspective, equation (\ref{massP}) is {\it exact } expansion of  the mass of 
the pseudoscalar meson $P$ in a infinite series of QCD operators 
$\langle \xi^n \rangle \sim \langle P(p) | \bar Q \gamma_{\nu} \gamma_5 (i z^{\sigma}  {\overset {\leftrightarrow} {D}}_{\sigma} )^n Q| 0 \rangle$
\beq
\frac {M_{P}^2} {4 M_Q^2} = \sum_{n=0}^{\infty} (n+1) \langle \xi^n  \rangle 
=1 + 3 \langle \xi^2 \rangle_A + 5 \langle \xi^4 \rangle_A + 7 \langle \xi^6 \rangle_A + O(\langle \xi^8 \rangle_A).
\label{expansion}
\eeq
Recall that equation (\ref{massP}) is valid at the scale when radiative corrections
are not very important. 

Formula (\ref{expansion}) has very simple interpretation. To understand what 
does it mean let us recall that the moments $\langle \xi^n \rangle_A$
has the meaning of the average of the n-th power relative velocity of quark-antiquark pair in the direction
of meson motion $\langle v_z^n \rangle$ \cite{Braguta:2006wr} \footnote{It is assumed that the meson is moving
along $z$-direction.}.  It is not difficult to understand that to get the average of 
the n-th power relative velocity $\langle v^n \rangle$,  the $\langle v_z^n \rangle$  must be multiplied by a factor of $n+1$, 
since to get average one should take the integral $\sim \int d \cos \theta  \cos^n \theta$. So, one has the 
series $1 + \langle v^2 \rangle + \langle v^4 \rangle + \langle v^6 \rangle+...$, which can be symbolically written
as
\beq
\frac {M_{P}^2} {4 M_Q^2} = \langle P(p) |  \frac 1 {1-\hat v^2}  | 0 \rangle.
\eeq
Evidently, this is quantum expression for the energy $E=2 M/\sqrt {1-v^2}$ of two relativistic particle
with zero net momentum moving with the speed $\hat v$.

In papers \cite{Braguta:2008tg, Braguta:2009xu, Braguta:2009df} it was 
pointed to the duality between light cone formalism and NRQCD 
description of hard exclusive processes. Although it was proved for some exclusive 
processes, today there is no strict proof of this duality. It would be interesting 
to know if it is possible to find formula in NRQCD dual to formula (\ref{massP}). To do this 
let us recall that in NRQCD one expands the mass of the pseudoscalar meson in operators
\beq
\langle v^n \rangle = \frac 1 {M_Q^2} \frac { \langle P(p) | \chi^+ ({ \bf {\overset {\leftrightarrow} {D}}} )^n \varphi | 0 \rangle} 
{\langle P(p) | \chi^+  \varphi | 0 \rangle},
\eeq
where $\varphi$ and $\chi$ are bispinors which create antiquark and destroy quark respectively.
To the leading order approximation one has 
$3 \langle \xi^2 \rangle = \langle v^2 \rangle$ \cite{Braguta:2006wr}. If one puts this equation to (\ref{expansion})
one gets
\beq
\frac {M_{P}^2} {4 M_Q^2} = 1+ \langle v^2 \rangle +O( \langle v^4 \rangle ).
\label{GK}
\eeq
which is well known Gremm-Kapustin relation \cite{Gremm:1997dq}. So, formulas 
(\ref{massP}), (\ref{expansion}) can be considered as an exact version 
of Gremm-Kapustin relation. 

Equation (\ref{massP}) can be used to test models of DAs. As an example let us take the models 
of 1S and 2S charmonia mesons proposed in papers \cite{Braguta:2006wr, Braguta:2007tq} 
at the central values of the parameters of these models. For the running mass of 
the c-quark $M_c$ the value $M_c=M_c^{\overline {MS}}(\mu=M_c^{\overline {MS}})=1.23$ 
will be taken \cite{Narison:1994ag}. Thus one 
gets $M_{\eta_c}=2.81$ GeV, $M_{\eta'_c}=3.53$ GeV, which is rather good estimation.

\section{Vector nonrelativistic mesons.}

In this section the formula similar to formula (\ref{massP}) but for the vector 
nonrelativistic meson will be derived. To do this one needs the following DAs
\beq
\langle V( p, \epsilon ) | \bar Q_{\alpha}(z)  Q_{\beta}(-z) | 0 \rangle = \frac {f_V M_V} 4
\int_0^1 d \xi e^{i (pz) \xi} \biggl \{ 
\biggl ( \hat {\epsilon} - \hat p \frac {({\epsilon}z)} {(pz)} \biggr ) \varphi_{\perp} (\xi) + 
\hat p \frac {({\epsilon}z)} {(pz)} \varphi_L(\xi) 
+  \nonumber \\ \frac {f_T} {f_V } \frac 1 {M_V}~ \sigma_{\mu \nu} {\epsilon}^{\mu} p^{\nu}~ \varphi_T(\xi,) + 
\frac 1 8 \biggl ( 1-  \frac {f_T} {f_V}  \frac{2 M_Q } { { M_V} }\biggr )~ 
e_{\mu \nu \sigma \rho} \gamma^{\mu} \gamma_5 {\epsilon}^{\nu} p^{\sigma} z^{\rho}~ \varphi_A(\xi)
\biggr \}_{\beta \alpha},
\label{vec}
\eeq
where the constants ${f_T}, {f_V}$ are defined as 
\beq
\langle V (p, \epsilon) | \bar Q \gamma_{\alpha} Q| 0 \rangle &=& f_{V} M_V \epsilon_{\alpha}, ~~~~~~~~~~
\langle V (p, \epsilon) | \bar Q \sigma_{\alpha \beta} Q| 0 \rangle = i f_{T}  
(p_{\alpha} \epsilon_{\beta} - p_{\beta} \epsilon_{\alpha} ).
\eeq
The DAs $\varphi_L(\xi), \varphi_T(\xi)$ are twist-2 DAs, the DAs $\varphi_{\perp}(\xi), \varphi_A(\xi)$ 
are twist-3 DAs.
The equations of motion for  DAs (\ref{vec}) give \cite{Ball:1998sk}
\beq
(n+1 ) \langle \xi^n \rangle_{\perp} =  \langle \xi^n \rangle_{L} + \frac {n(n-1)} {2} (1-\delta) 
\langle \xi^{n-2} \rangle_A,  \\ \nonumber
\frac 1 2 (n+2 ) (1-\delta ) \langle \xi^n \rangle_A =  \langle \xi^n \rangle_{\perp} -
\delta \langle \xi^{n} \rangle_T,
\label{eqV}
\eeq
where $\langle \xi^n \rangle_{L,T,\perp, A}$ are the moments of the DAs 
$\varphi_L (\xi), \varphi_T(\xi), \varphi_{\perp}(\xi), \varphi_A(\xi)$, 
$\delta(\mu)=2 f_T/f_V M_Q /M_V$. 

Equations (\ref{eqV}) can be exactly solved \cite{Ball:1998sk}. For instance, 
the solution for the DA $\varphi_A$ is
\beq
\varphi_A (\xi) = C_0 + C_1 \xi + \frac 1 {1- \delta} 
\biggl ( 
(1+\xi) \int_{\xi}^1 \frac {\psi(t)} {1+t} dt + (1-\xi) \int_{-1}^{\xi} \frac {\psi(t)} {1-t} dt
\biggr ),
\eeq
where $\psi(\xi) = \varphi_L (\xi) + \delta \xi \varphi'_T (\xi)$. The function 
$\varphi_A(\xi)$ must be even, normalized and it must be suppressed in the end point region.
These requirements are fulfilled if $C_0=C_1=0$ and $\delta$ is equal to
\beq
\frac  {f_T } {f_V} \frac {2 M_Q} {M_V} = \frac { \int_{-1}^1 \frac {d \xi} {1-\xi^2 } 
\varphi_L (\xi) } {\int_{-1}^1  {d \xi} \frac {1+\xi^2} {(1-\xi^2)^2 } 
\varphi_T (\xi) }.
\eeq
This formula is the analog of equation (\ref{massP}) but for vector meson $V$. It 
is more complicated than equation (\ref{massP}) since vector 
meson has two independent constants $f_T$ and $f_V$. Note that if one expands 
now right hand side of the last equation and puts Gremm-Kapustin relation
(\ref{GK}) one will get NRQCD expression for the ratio $f_T/f_V$ \cite{Braaten:1998au, Braguta:2007ge}
\beq
\frac {f_T} {f_V}=1-\frac {\langle v^2 \rangle} {6}
\eeq

\begin{acknowledgments}
The author thanks A.K. Likhoded, A.V. Luchinsky for useful discussion.
This work was partially supported by Russian Foundation of Basic Research under grant 07-02-00417, 
by president grant MK-140.2009.2 and scientific school grant SS-679.2008.2.
\end{acknowledgments}


\begin{thebibliography}{**}

% introduction


%\cite{Lepage:1980fj}
\bibitem{Lepage:1980fj}
  G.~P.~Lepage and S.~J.~Brodsky,
  %``Exclusive Processes In Perturbative Quantum Chromodynamics,''
  Phys.\ Rev.\ D {\bf 22}, 2157 (1980).
  %%CITATION = PHRVA,D22,2157;%%

%\cite{Chernyak:1983ej}
\bibitem{Chernyak:1983ej}
  V.~L.~Chernyak and A.~R.~Zhitnitsky,
  %``Asymptotic Behavior Of Exclusive Processes In QCD,''
  Phys.\ Rept.\  {\bf 112}, 173 (1984).
  %%CITATION = PRPLC,112,173;%%




%\cite{Bodwin:2006dm}
\bibitem{Bodwin:2006dm}
  G.~T.~Bodwin, D.~Kang and J.~Lee,
  %``Reconciling the light-cone and NRQCD approaches to calculating e+ e- -->
  %J/psi + eta/c,''
  Phys.\ Rev.\  D {\bf 74}, 114028 (2006)
  [arXiv:hep-ph/0603185].
  %%CITATION = PHRVA,D74,114028;%%


%\cite{Ma:2006hc}
\bibitem{Ma:2006hc}
  J.~P.~Ma and Z.~G.~Si,
  %``NRQCD factorization for twist-2 light-cone wave-functions of charmonia,''
  Phys.\ Lett.\  B {\bf 647}, 419 (2007)
  [arXiv:hep-ph/0608221].
  %%CITATION = PHLTA,B647,419;%%




%\cite{Braguta:2006wr}
\bibitem{Braguta:2006wr}
  V.~V.~Braguta, A.~K.~Likhoded and A.~V.~Luchinsky,
  %``The study of leading twist light cone wave function of eta/c meson,''
  Phys.\ Lett.\  B {\bf 646}, 80 (2007)
  [arXiv:hep-ph/0611021].
  %%CITATION = PHLTA,B646,80;%%

%\cite{Braguta:2007fh}
\bibitem{Braguta:2007fh}
  V.~V.~Braguta,
  %``The study of leading twist light cone wave functions of J/psi meson,''
  Phys.\ Rev.\  D {\bf 75}, 094016 (2007)
  [arXiv:hep-ph/0701234].
  %%CITATION = PHRVA,D75,094016;%%



%\cite{Braguta:2007tq}
\bibitem{Braguta:2007tq}
  V.~V.~Braguta,
  %``The study of leading twist light cone wave functions of 2S state charmonium
  %mesons,''
  Phys.\ Rev.\  D {\bf 77}, 034026 (2008)
  [arXiv:0709.3885 [hep-ph]].
  %%CITATION = PHRVA,D77,034026;%%



%\cite{Choi:2007ze}
\bibitem{Choi:2007ze}
  H.~M.~Choi and C.~R.~Ji,
  %``Perturbative QCD analysis of exclusive $J/\psi+\eta_c$ production in
  %$e^+e^-$ annihilation,''
  Phys.\ Rev.\  D {\bf 76}, 094010 (2007)
  [arXiv:0707.1173 [hep-ph]].
  %%CITATION = PHRVA,D76,094010;%%


%\cite{Feldmann:2007id}
\bibitem{Feldmann:2007id}
  T.~Feldmann and G.~Bell,
  %``Light-Cone Distribution Amplitudes for Non-Relativistic Bound States,''
  AIP Conf.\ Proc.\  {\bf 964}, 110 (2007)
  [arXiv:0711.4014 [hep-ph]].
  %%CITATION = APCPC,964,110;%%

%\cite{Bell:2008er}
\bibitem{Bell:2008er}
  G.~Bell and T.~Feldmann,
  %``Modelling light-cone distribution amplitudes from non-relativistic bound
  %states,''
  JHEP {\bf 0804}, 061 (2008)
  [arXiv:0802.2221 [hep-ph]].
  %%CITATION = JHEPA,0804,061;%%

%\cite{Braguta:2008qe}
\bibitem{Braguta:2008qe}
  V.~V.~Braguta, A.~K.~Likhoded and A.~V.~Luchinsky,
  %``Leading twist distribution amplitudes of P-wave nonrelativistic mesons,''
  Phys.\ Rev.\  D {\bf 79}, 074004 (2009)
  [arXiv:0810.3607 [hep-ph]].
  %%CITATION = PHRVA,D79,074004;%%

%\cite{Hwang:2008qi}
\bibitem{Hwang:2008qi}
  C.~W.~Hwang,
  %``Study of quark distribution amplitudes of 1S and 2S heavy quarkonium
  %states,''
  Eur.\ Phys.\ J.\  C {\bf 62}, 499 (2009)
  [arXiv:0811.0648 [hep-ph]].
  %%CITATION = EPHJA,C62,499;%%



%\cite{Hwang:2009cu}
\bibitem{Hwang:2009cu}
  C.~W.~Hwang,
  %``Leading-twist light cone distribution amplitudes for p-wave heavy
  %quarkonium states,''
  JHEP {\bf 0910}, 074 (2009)
  [arXiv:0906.4412 [hep-ph]].
  %%CITATION = JHEPA,0910,074;%%




%\cite{Ball:1998je}
\bibitem{Ball:1998je}
  P.~Ball,
  %``Theoretical update of pseudoscalar meson distribution amplitudes of  higher
  %twist: The nonsinglet case,''
  JHEP {\bf 9901}, 010 (1999)
  [arXiv:hep-ph/9812375].
  %%CITATION = JHEPA,9901,010;%%


%\cite{Ball:1998sk}
\bibitem{Ball:1998sk}
  P.~Ball, V.~M.~Braun, Y.~Koike and K.~Tanaka,
  %``Higher twist distribution amplitudes of vector mesons in {QCD}: Formalism
  %and twist three distributions,''
  Nucl.\ Phys.\  B {\bf 529}, 323 (1998)
  [arXiv:hep-ph/9802299].
  %%CITATION = NUPHA,B529,323;%%
  
%\cite{Bodwin:1994jh}
\bibitem{Bodwin:1994jh}
  G.~T.~Bodwin, E.~Braaten and G.~P.~Lepage,
  % ``Rigorous QCD analysis of inclusive annihilation and production of heavy
  %quarkonium,''
  Phys.\ Rev.\ D {\bf 51}, 1125 (1995)
  [Erratum-ibid.\ D {\bf 55}, 5853 (1997)]
  [arXiv:hep-ph/9407339].
  %%CITATION = HEP-PH 9407339;%%


%\cite{Braguta:2008tg}
\bibitem{Braguta:2008tg}
  V.~V.~Braguta,
  %``Double charmonium production at B-factories within light cone formalism,''
  PoS C {\bf ONFINEMENT8}, 097 (2008)
  [Phys.\ Rev.\  D {\bf 79}, 074018 (2009)]
  [arXiv:0811.2640 [hep-ph]].
  %%CITATION = PHRVA,D79,074018;%%



%\cite{Braguta:2009xu}
\bibitem{Braguta:2009xu}
  V.~V.~Braguta and V.~G.~Kartvelishvili,
  %``Decay eta_b->J/\psi J/\psi in light cone formalism,''
  arXiv:0907.2772 [hep-ph].
  %%CITATION = ARXIV:0907.2772;%%








%\cite{Braguta:2009df}
\bibitem{Braguta:2009df}
  V.~V.~Braguta, A.~K.~Likhoded and A.~V.~Luchinsky,
  %``Double charmonium production in exclusive bottomonia decays,''
  Phys.\ Rev.\  D {\bf 80}, 094008 (2009)
  [arXiv:0902.0459 [hep-ph]].
  %%CITATION = PHRVA,D80,094008;%%


%\cite{Gremm:1997dq}
\bibitem{Gremm:1997dq}
  M.~Gremm and A.~Kapustin,
  %``Annihilation of S-wave quarkonia and the measurement of alpha(s),''
  Phys.\ Lett.\  B {\bf 407}, 323 (1997)
  [arXiv:hep-ph/9701353].
  %%CITATION = PHLTA,B407,323;%%

%\cite{Narison:1994ag}
\bibitem{Narison:1994ag}
  S.~Narison,
  %``A Fresh Look Into The Heavy Quark Mass Values,''
  Phys.\ Lett.\  B {\bf 341}, 73 (1994)
  [arXiv:hep-ph/9408376].
  %%CITATION = PHLTA,B341,73;%%

%\cite{Braaten:1998au}
\bibitem{Braaten:1998au}
  E.~Braaten and Y.~Q.~Chen,
  %``Renormalons in electromagnetic annihilation decays of quarkonium,''
  Phys.\ Rev.\  D {\bf 57}, 4236 (1998)
  [Erratum-ibid.\  D {\bf 59}, 079901 (1999)]
  [arXiv:hep-ph/9710357].
  %%CITATION = PHRVA,D57,4236;%%

%\cite{Braguta:2007ge}
\bibitem{Braguta:2007ge}
  V.~V.~Braguta,
  %``The study of double vector charmonium mesons production at B-factories
  %within light cone formalism,''
  Phys.\ Rev.\  D {\bf 78}, 054025 (2008)
  [arXiv:0712.1475 [hep-ph]].
  %%CITATION = PHRVA,D78,054025;%%




\end{thebibliography}
\end{document}